# Real-fluid simulation of ammonia cavitation in a heavy-duty fuel injector

Hesham Gaballa[1], Chaouki Habchi*[1], Jean-Charles De Hemptinne[1], Gerard Mouokue [2]
[1]IFP Energies Nouvelles, Institut Carnot Transports Energies, 1 et 4 Avenue de Bois-Préau, 92852 Rueil-Malmaison, France
[2]Woodward L'Orange GmbH, Porschestraße 8, 70435 Stuttgart, Germany
*Corresponding author: chawki.habchi@ifpen.fr

**Abstract**
The reduction of greenhouse gases (GHG) emitted into the earth's atmosphere, such as carbon dioxide, has obviously become a priority. Replacing fossil fuels with cleaner renewable fuels (such as ammonia) in internal combustion engines for heavy-duty vehicles is one promising solution to reduce GHG emissions. This paper aims to study the cavitation formation in a heavy-duty injector using ammonia as fuel. The simulation is carried out using a fully compressible two-phase multi-component real-fluid model (RFM) developed in the CONVERGE CFD solver. In the RFM model, the thermodynamic and transport properties are stored in a table which is used during the run-time. The thermodynamic table is generated using the in-house Carnot thermodynamic library based on vapor-liquid equilibrium calculations coupled with a real-fluid equation of state. The RFM model allows to consider the effects of the dissolved non-condensable gas such as nitrogen on the phase change process. The obtained numerical results have confirmed that the model can tackle the phase transition phenomenon under the considered conditions. In contrast to previous numerical studies of the cavitation phenomenon using hydrocarbon fuels, the formed cavitation pockets were found to be primarily composed of ammonia vapor due to its high vapor pressure, with minimal contribution of the dissolved non-condensable nitrogen.

**1. Introduction**
The recent stringent emissions legislation poses new challenges to the continuous use of diesel-powered internal combustion engines (ICE) due to their carbon dioxide emissions and urban pollution, which accelerate climate change and are linked to severe health problems, respectively. The decarbonisation of the diesel engine can be achieved by the implementation of alternative fuels such as ammonia, due to its carbon-free structure, its storage and transportation safety, and reasonable production cost [1]. To this goal, further understanding of the physical phenomena that take place inside the fuel injector such as cavitation is indeed essential for further development of such ammonia powered ICE.
Various numerical investigations of the cavitation phenomenon can be found in the literature for hydrocarbon fuels. The numerical models were based on the Eulerian-Lagrangian [2] or Eulerian-Eulerian [3-6] approaches. The phase change process is commonly modelled using non-equilibrium mass transfer models [6] or equilibrium mass transfer models [3-5] assuming thermodynamic equilibrium. In addition, cavitation models usually consider the liquid, vapor, and ambient gas, excluding the dissolved gas in the liquid phase. However, recent studies [3-6] have highlighted the effect of dissolved non-condensable gas on the cavitation development. Indeed, two cavitation regimes have been identified, namely vaporous cavitation and gaseous cavitation. On the one hand, vaporous cavitation takes places, when the local pressure decreases to or below the liquid (including dissolved gas) saturation pressure. On the other hand, gaseous cavitation is formed, due to the expansion of non-condensable gas bubbles, as the pressure drops, but not necessarily below the liquid saturation pressure (see [5] for a detailed discussion).
The cavitation simulation considering the real-fluid effects has been recently shown in [4] using the Peng-Robinson (PR) EoS [18] and vapor-liquid equilibrium (VLE) calculations. However, one issue reported by the authors is the model computational efficiency, as thermodynamic solver calculations were expensive. Accordingly, a more robust and efficient tabulation approach could be one remedy to the computationally demanding VLE solver. For instance, Vidal et al. [11] employed a tabulated thermodynamic approach based on the PC-SAFT EoS and VLE calculations to investigate the cavitation formation of a multi-component diesel fuel in a high-pressure fuel injector. However, they have not considered the dissolved non-condensable gas in their study.
Accordingly, the current work proposes a fully compressible multi-component two-phase real-fluid model (RFM) [9,10] closed by a thermodynamic equilibrium tabulation method based on a real-fluid EoS. The main goal is to investigate the cavitation formation in a heavy-duty injector using ammonia as a fuel. The proposed RFM model does not have fitting coefficients for modelling cavitation formation and collapse. Besides, it allows to consider the real-fluid thermo-transport properties of the fuel-non-condensable gas mixture. In addition, the separate contribution of vaporous and gaseous cavitation to the total void formation can be quantified. This paper is organized as follows. Section 2 describes the proposed RFM model. Then, the numerical setup and the simulation results of cavitation in a heavy-duty diesel injector with ammonia as fuel are discussed in Section 3. Finally, the conclusions are presented in section 4.

## 2. The real-fluid model (RFM)
### 2.1 Governing equations
The diffused interface two-phase flow model adopted in the current work is a four-equation model that is fully compressible and considers multi-component is both phases under the assumptions of thermal and mechanical equilibrium as follows,

$$\frac{\partial \rho}{\partial t} + \frac{\partial \rho u_i}{\partial x_i} = 0 \tag{1}$$

$$\frac{\partial \rho u_i}{\partial t} + \frac{\partial \rho u_i u_j}{\partial x_j} = -\frac{\partial P}{\partial x_i} + \frac{\partial \tau_{ij}}{\partial x_j}, \quad \tau_{ij} = \mu \left( \frac{\partial u_i}{\partial x_j} + \frac{\partial u_j}{\partial x_i} - \frac{2}{3} \frac{\partial u_k}{\partial x_k} \delta_{ij} \right) \tag{2}$$

$$\frac{\partial \rho e}{\partial t} + \frac{\partial \rho e u_j}{\partial x_j} = -P \frac{\partial u_j}{\partial x_j} + \tau_{ij} \frac{\partial u_i}{\partial x_j} + \frac{\partial}{\partial x_j} \left( \lambda \frac{\partial T}{\partial x_j} \right) + \frac{\partial}{\partial x_j} \left( \rho D \sum_k h_k \frac{\partial Y_k}{\partial x_j} \right) \tag{3}$$

$$\frac{\partial \rho Y_k}{\partial t} + \frac{\partial \rho Y_k u_j}{\partial x_j} = \frac{\partial}{\partial x_j} \left( \rho D \frac{\partial Y_k}{\partial x_j} \right) \tag{4}$$

where ($\tau_{ij}$) is the viscous stress tensor, ($\rho$) is the density, ($u_i$) is the velocity, ($P$) is the pressure, ($e$) is the specific internal energy, ($T$) is the temperature. Besides, ($Y_k, h_k$) are the mass fraction and specific enthalpy of species $k$, respectively. The thermal conductivity ($\lambda$) and the dynamic viscosity ($\mu$) covers laminar and turbulent contributions. The laminar contribution of ($\lambda, \mu$) is computed by Chung et al. [12] correlations. The turbulent conductivity is calculated using a given turbulent Prandtl number and the turbulent viscosity is given by the adopted turbulence model. Finally, the laminar and turbulent mass diffusion coefficients are estimated using a given Schmidt number.

### 2.2 Tabulated thermodynamic closure
The fully compressible multi-component two-phase flow system described above is closed by a tabulated real-fluid EoS adopting a local thermodynamic equilibrium hypothesis. To consider the two-phase properties as well as the phase change phenomenon, the EoS is not solely sufficient, and thus vapor-liquid equilibrium calculations are also included in the RFM model. The current work employs a pre-tabulation approach, where before the CFD simulation, a uniform thermodynamic table is generated using the IFPEN-Carnot thermodynamic library. The thermodynamic library performs the VLE calculation using a robust isothermal-isobaric (TP) flash [13] coupled to a real-fluid EoS. The tabulated properties include the thermodynamic equilibrium density, internal energy, thermodynamic derivatives as heat capacity, sound speed, and transport properties. The thermodynamic table axes are the temperature ($T$), pressure ($P$) or the decimal logarithm of pressure($\log_{10} P$), and species mass fraction ($Y_k, k = 1, N_s - 1$), where ($N_s$) is the total number of species. In the current study, the ($\log_{10} P$), approach is employed, as it can provide sufficient resolution for the pressure axis to capture the cavitation formation with adequate number of points compared to a linear ($P$) axis. During the simulation, the required tabulated quantities are interpolated using the Inverse Distance Weighting (IDW) method [14]. The thermodynamic table is coupled with the CONVERGE CFD solver [15], as detailed in previous studies [9,10]. The adopted tabulation approach is one remedy to the direct evaluation of the costly phase equilibrium solver during the CFD simulation, especially when coupled with a complex real-fluid EoS [7,8]. In this study, the thermodynamic closure of the (Ammonia/nitrogen) binary mixture is based on a volume translated Peng-Robinson (VTPR)-EoS with a uniform thermodynamic table resolution in (T, $\log_{10} P$, $Y_{NH_3}$), axes of (201×201×21) points covering ranges of (200-600 K, 0.001-760 bar, 0.9999-1) corresponding to the case setup described in the next section.

## 3. Cavitation simulation results and discussion
### 3.1 Test case setup
The investigated test case configuration is based on a common rail 7-hole heavy-duty diesel injector of Woodward L'Orange GmbH. The simulated geometry here considered only one seventh of the full injector geometry, as shown in Fig. 1. The needle valve was assumed to be still at its maximum lift of 480 μm. The operating conditions of the injector are summarized in Table 1. The base mesh size is set to 100 μm, where various mesh embedding levels have been employed, resulting in three grids (see Table. 2) used for grid sensitivity analysis. The employed boundary conditions are described in Fig. 1. A pressure and temperature boundary conditions of (750 bar, 343 K) are imposed at the injector inlet. A hemispherical volume is added to the orifice exit to avoid the interference of the outlet boundary on the cavitation development inside the orifice. A pressure outlet boundary condition is set at the outlet volume, which is relaxed from 750 bar to 50 bar in a time interval of 0.1 ms to facilitate the simulation start-up. The entire computational domain including the outlet hemispherical volume is initialized with pressure of 750 bar and temperature of 343 K. The initial phase state is single liquid phase $(\alpha_l = 1)$ with an initial amount of dissolved nitrogen of $(Y_{N_2} = 2e-05)$. The mass fraction of dissolved nitrogen in the feed is chosen to be smaller than the saturation value $(Y_{N_2} \approx 1e-04)$ estimated from the VLE solver at (T=298 K, P=10.3 bar), assuming that the liquid ammonia is stored at these conditions, as reported in [1].

**Table 1.** Operating conditions of the injector

| $P_{inj}(bar)$ | $T_{inj}(K)$ | $P_{out}(bar)$ |
|---|---|---|
| 750 | 343 | 50 |

**Table 2.** Grid resolutions employed in the grid sensitivity study.

| Grid no. | Min. cell size (µm) | No. of cells (M) |
|---|---|---|
| 1 | 12.5 | 0.83 |
| 2 | 6.25 | 1.15 |
| 3 | 3.125 | 2.84 |

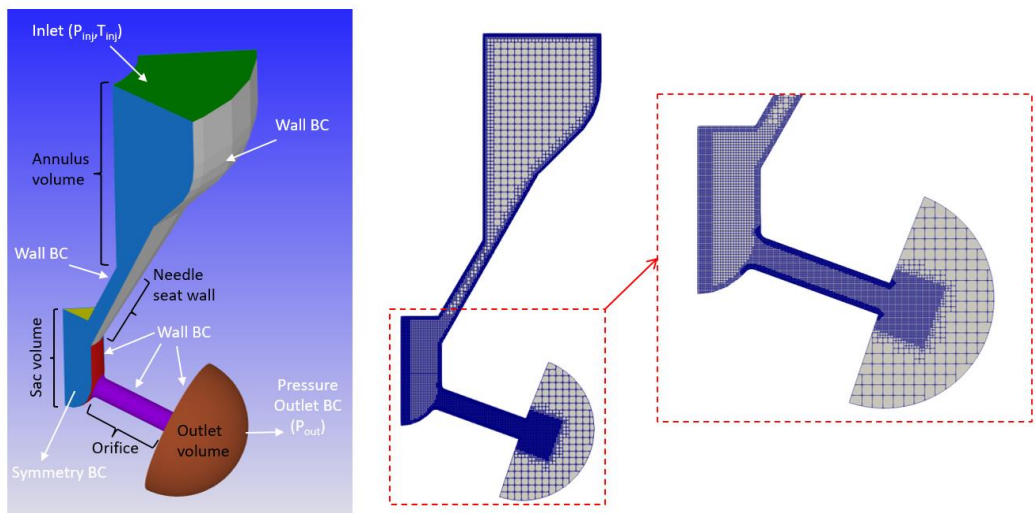

**Figure 1**. Simulated configuration (only 1/7 of the full injector geometry) along with the boundary conditions. The computational mesh is also shown at the domain central cut section. The insert shows the refined mesh within the injector orifice.

URANS simulations are carried out using the RNG $k-\epsilon$ turbulence model [16] with the default model constants and standard wall treatment. Simulations are carried out using the RFM model implemented in CONVERGE V3.1.6 employing an updated real-fluid PISO algorithm, which is consistent with the employed real-fluid EoS. The spatial discretization is second-order accurate using a central difference scheme. The time integration is achieved by a second-order Crank-Nicolson scheme for the momentum equation and a first-order implicit Euler scheme for the rest of the equations. The time step is around 0.6-0.7 ns and adjusted automatically based on a maximum CFL number of 0.5.

### 3.2 Numerical results
First, to study the effect of the mesh size on the simulation predictions, Fig. 2 shows the variation of the velocity coefficient $(C_v)$ as function of time for the different grid resolutions. A slight variation of the $(C_v)$ can be observed between Grid 2 and Grid 3, showing that grid convergence is fairly achieved. Accordingly, Grid 2 with a minimum mesh resolution of 6.25 µm at the orifice wall is used for further calculations and the associated results will be further discussed.

Figure 3 illustrates the temporal evolution of the gas volume fraction $(\alpha_g)$ at the orifice mid-plane. It shows that the cavitation incepts (t=0.075 ms) near the upper side of the orifice inlet. The generated cavitation zone further increases with time and extends to fill a bigger region near the upper wall of the orifice. Besides, at (t=0.095 ms), a cavity can be observed near the central region of the orifice, which is further transported to the orifice exit as time elapsed. At the time greater than (t=0.155 ms), the cavitation pockets show very little variation, indicating that the in-nozzle flow has reached a quasi-steady state condition.

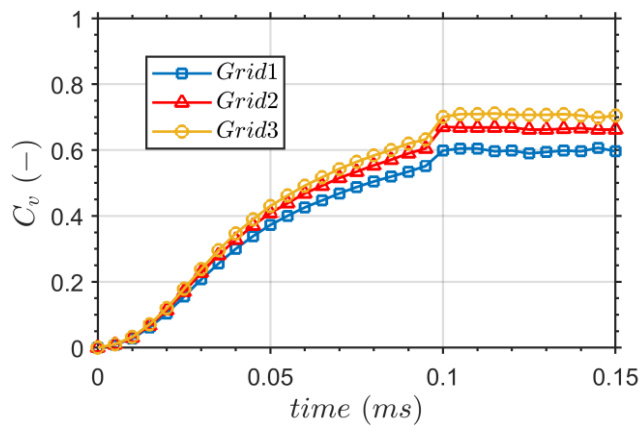

**Figure 2.** Temporal evolution of the of the velocity coefficient ($C_v$) for the different grid resolutions.

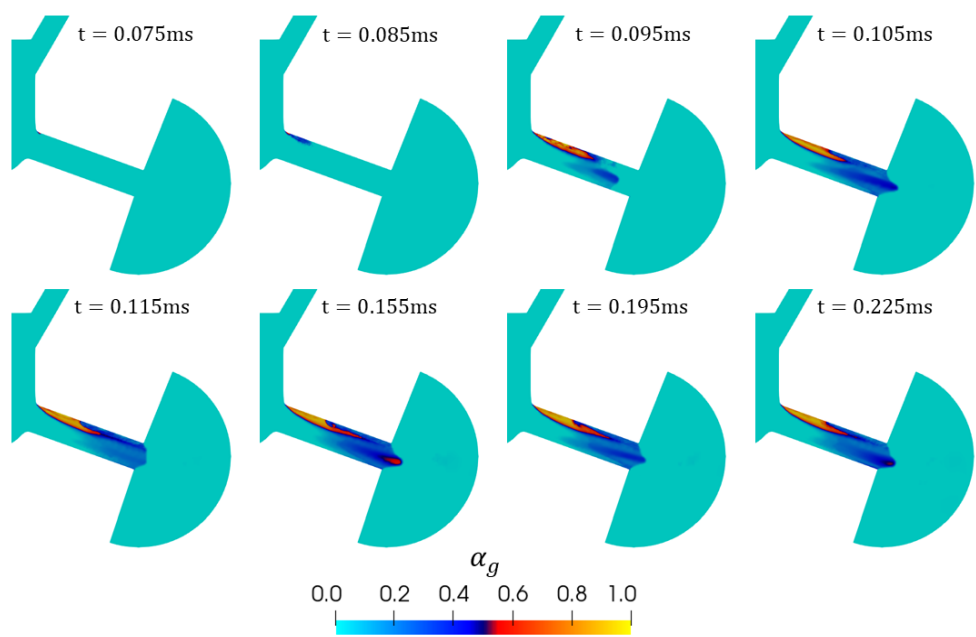

**Figure 3**. Temporal evolution the gas volume fraction ($\alpha_g$) at the mid-plane of the injector.

To further visualize the cavitation formation within the injector orifice, Fig. 4 shows the gas volume fraction and velocity magnitude distributions at different slices along the orifice cross-section at (t=0.2 ms), where the flow has reached a quasi-steady state condition. It can be seen the cavitation regions that is formed mainly near the upper wall of the orifice. These regions also correspond to locations, where the velocity is reduced compared to that at the orifice centre (Fig. 4 right). Indeed, as the flow arrives to the orifice inlet, it could not accommodate for the abrupt change in direction, leading to flow separation and creating a low-pressure region, where cavitation starts to develop. In addition, it can be also observed that cavitation regions are also developed near the orifice centre (Fig. 4 left). The reason for these cavitation regions will be specified later in the discussion.

It is also of interest to quantify the contribution of both the fuel ($NH_3$) and dissolved non-condensable gas ($N_2$) to the formed gaseous phase. The employed VLE based model can provide valuable information regarding the phase change process and the composition of each species in each phase. Indeed, Fig. 5. illustrates the temporal variation of gas cavities with the iso-surface ($\alpha_g = 0.9$) colored by the volume fraction of nitrogen in the gas phase ($\alpha_{g,N_2}$) and the pressure. The ($\alpha_{g,N_2}$) is defined as ($\alpha_{g,N_2} = y_{N_2}\alpha_g$), where ($y_{N_2}$) is the mole fraction of nitrogen in the gas phase. The results show that contribution of the dissolved $N_2$ to the cavitation pockets is very small (see the palette of Fig. 5 top). Based on the constraint ($y_{N_2} + y_{NH_3} = 1$), it can be concluded that the cavitation pockets are mainly dominated by the ammonia vapor. In other words, vaporous cavitation is the dominant phase transition process compared to gaseous cavitation. This could be attributed to the high saturation pressure (33 bar at 343K) of ammonia, which facilitates the formation of the vaporous cavitation. Indeed, the pressure distribution (Fig. 5 bottom) shows that the pressure at the gas cavities is much lower than the saturation pressure, leading to high contribution of ammonia vapor to the gas phase compared to that of nitrogen. Such behaviour of ammonia is different from that reported in previous cavitation studies [4-6,17] of hydrocarbon fuels, where gaseous cavitation was found to have a significant contribution to the formed gas cavities.

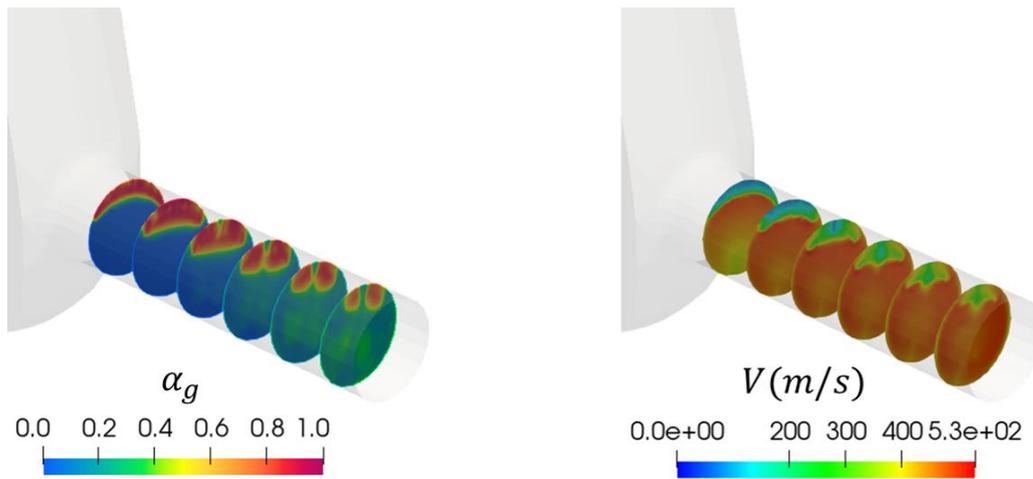

**Figure 4.** Distributions of the gas volume fraction (left) and velocity magnitude (right) at different slices along the orifice cross-section at t=0.2 ms.

For further analysis of the cavitation development inside the orifice, the pressure and gas volume fraction variation along the orifice length at both the top wall (r/R=0.99) and center line (r/R=0) are depicted in Fig. 6. On the one hand, it can be observed that the pressure exhibits a sharp decrease at the top wall slightly after the orifice inlet. As the pressure drops below the saturation pressure (dotted line in Fig. 6 left), cavitation starts to form as shown by the increase of the ($\alpha_g$) (Fig. 6 center).

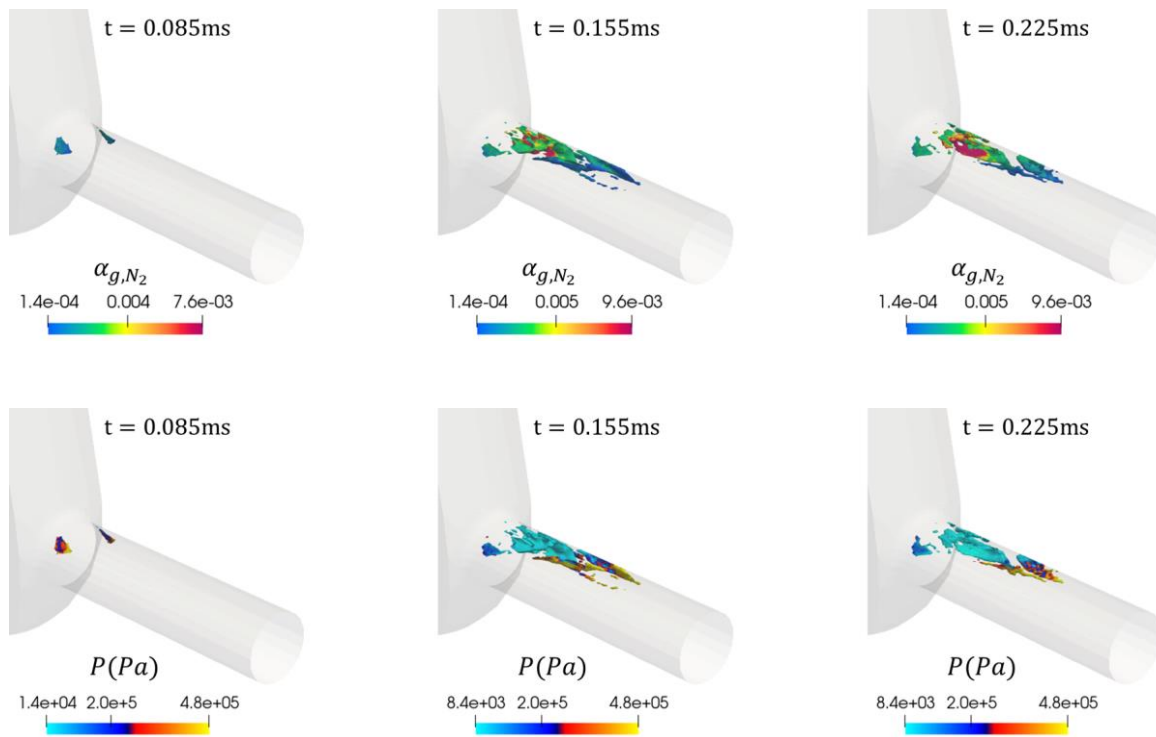

**Figure 5.** Temporal evolution of the iso-surface ($\alpha_g = 0.9$) coloured by the volume fraction of nitrogen in the gas-phase (top) and the pressure (bottom).

On the other hand, at the centre of the orifice, a smother pressure drop takes place. Indeed, the pressure decreases below the saturation pressure at a longer distance inside the orifice compared to the top wall. Such pressure drop below the saturation value at the orifice centre explains the formed gaseous region that have been observed near the orifice centre (see Fig. 6 right). However, the amount of gas generated at the orifice centre is much lower compared to that at the top wall near the orifice inlet, as depicted by the gas volume fraction variation in Fig. 6.

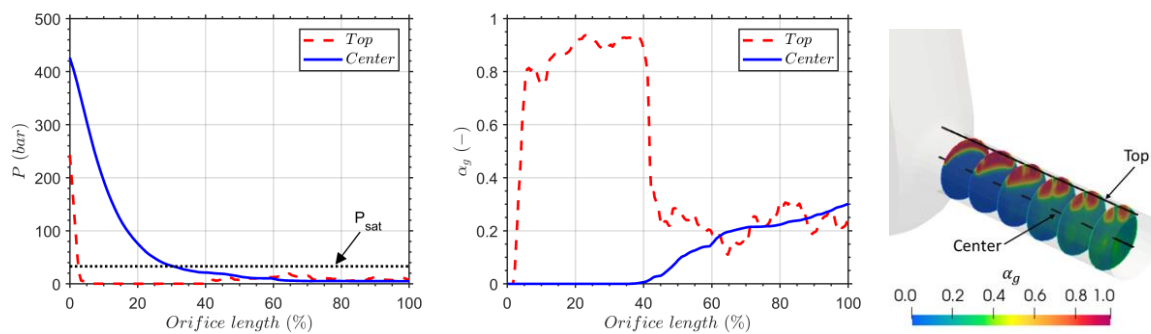

**Figure 6**. Variation of the pressure (left) and the total gas volume fraction (center) along the orifice length for the locations shown in (right) at t=0.2 ms. The ($P_{sat}$) denotes the saturation pressure.

## 4. Conclusion

In this paper, the cavitation formation in a heavy-duty injector has been numerically investigated using ammonia as a fuel. Simulations are carried out using the proposed real-fluid model (RFM) in this study. The main conclusions can be summarized as follows,

- The obtained numerical results have shown that the RFM model is able to dynamically predict the phase transition process under the considered industrial injector configuration.
- The formed cavitation pockets inside the injector orifice are revealed to be mainly composed of ammonia vapor, with a slight contribution of the dissolved non-condensable nitrogen. Such behaviour could be attributed to the relatively high saturation pressure of ammonia. For instance, $P_{sat}$=33 bar at T=343K.
- The formation of gaseous cavities is found to take place not only near the upper wall of the orifice, but also near the orifice centre, as the pressure deceases below the saturation pressure. However, the amount of gas generated at the orifice centre is much lower compared to that at the top wall near the orifice inlet.
- The employed thermodynamic table based on ($\log_{10} P$) has also shown its effectiveness to provide adequate table resolution for the pressure axis, which is required for cavitation simulations, as the phase change is mainly driven by the pressure variation.

**Acknowledgements**

This project has received funding from the European Union Horizon 2020 Research and Innovation programme. Grant Agreement No 861002 for the EDEM project.